\documentstyle[aps,prl,floats,epsfig]{revtex}

\newcommand{\eins}{\mbox{$1 \hspace{-1.0mm}  {\bf l}$}}
\newcommand{\one}{\mbox{$1 \hspace{-1.0mm}  {\bf l}$}}
\newcommand{\be}{\begin{equation}}
\newcommand{\ee}{\end{equation}}
\newcommand{\bea}{\begin{eqnarray}}
\newcommand{\eea}{\end{eqnarray}}
\newcommand{\half}{\mbox{$\textstyle \frac{1}{2}$}}
\newcommand{\quarter}{\mbox{$\textstyle \frac{1}{4}$}}
\newcommand{\shalf}{\mbox{$\textstyle \frac{1}{\sqrt{2}}$}}
\newcommand{\sthird}{\mbox{$\textstyle \frac{1}{\sqrt{3}}$}}

\newcommand{\ket}[1]{ | \, #1  \rangle}
\newcommand{\bra}[1]{ \langle #1 \,  |}
\newcommand{\braket}[2]{\left< #1 \right| #2 \rangle}
\newcommand{\proj}[1]{\ket{#1}\bra{#1}}

\newcommand{\mytext}[1]{\mbox{ #1}}

\newcommand{\calw}{\mbox{$\cal W$}}
\newcommand{\calh}{\mbox{$\cal H$}}

\begin{document} \draft

\title{Characterizing Entanglement}

\author{
Dagmar Bru\ss \\
{\em Institut f\"ur Theoretische Physik, 
Universit\"at Hannover\\
Appelstr. 2, D-30167 Hannover, Germany}\\
{\small email: bruss@itp.uni-hannover.de}
}
\date{Received \today}
\maketitle

\begin{abstract}
Quantum entanglement is at the heart of many tasks in quantum
information. Apart from simple cases (low dimensions, few particles,
pure states), however, the mathematical structure of entanglement 
is not yet fully understood. This tutorial is an introduction to our 
present knowledge about 
how to decide whether a given state is separable or entangled, 
how to characterize entanglement via 
witness operators, how to classify entangled states according to their 
usefulness (i.e. distillability), and
 how to quantify entanglement with appropriate 
measures.
\end{abstract}

\section{Introduction}

The role of entanglement in quantum information
 processing is manifold. Rather than considering entanglement
 as a mystery, like in the early years of quantum mechanics, 
  it is nowadays viewed as a resource for certain
 tasks that can be performed faster or in a more secure way 
 than classically. This genuinely new aspect of quantum properties
 has launched intensive experimental efforts 
  to create entangled states  and theoretical efforts to
 understand the mathematical structure of entanglement. 
 This tutorial  presents the  status regarding our 
  understanding of entanglement. 

As the number of articles about entanglement has ernomously 
increased during the last ten years, it is almost impossible to 
give a complete overview, and this is not the purpose of this
article. It will rather introduce the reader to  the
established
knowledge and some important tools in this field, and 
discuss  some of the questions that remain open at present.

Throughout this tutorial, we will mostly consider entanglement
of just two parties, unless stated 
otherwise explicitly. Most concepts
can be explained best with bipartite systems; some of them
could then be generalized to more parties in an evident way,
for others the situation changes completely
 for more than two parties. Not many results
are known for multipartite systems. In section
\ref{many} some reasons for this will become clear from
studying tripartite states.

Various aspects of entanglement have recently been summarized in the 
following review articles,  which were partially used 
as a source for this tutorial: 
the ``primer'' \cite{primer}
aims at introducing the non-expert reader 
to the problem of separability and distillability of quantum states.
The Horodecki family discusses entanglement
in the context of quantum communication \cite{hororev}, where 
the distillability properties of a given state are important. 
B. Terhal summarizes the use of witness operators for detecting
entanglement in \cite{terhal}. Various entanglement measures
are presented in the context of the theorem of their uniqueness
in \cite{mdonald}.
Other reviews on theoretical and experimental aspects of entanglement
can be found in the first issues of the newly launched journal QIC
\cite{qic}.

This paper is organized as follows: Section \ref{whatis} presents various
possible answers to the question  ``what  is entanglement?'',
thus shedding light upon different facets of quantum correlations.
In section \ref{seporent} several criteria are introduced that
allow to distinguish separable from entangled states.
Section \ref{useful} discusses the possibility to distill entanglement,
and gives a distillability criterion. Finally, section 
\ref{howmuch} concerns attempts to quantify entanglement via 
entanglement measures. Some important measures are defined and their
properties are discussed. The classification of entangled states
according to their Schmidt number is introduced, and a
generalization to tripartite states is included.

\section{ What {\em is} entanglement?}
\label{whatis}

It is nearly 70 years ago that  Erwin Schr\"odinger 
 gave the  name 
``Verschr\"ankung'' to  a correlation of
quantum nature 
\cite{es}. In colloquial German for non-physicists
this term is only used in
the sense of ``folding the arms''. 
It was then rather loosely translated to 
``entanglement'', with more inspiring connotations.

Over the decades the meaning of the word ``entanglement'' has 
changed its flavour. The following list is an attempt to 
sketch 
the attitude towards entanglement of various important
persons in the fields of foundations of quantum physics and 
later in quantum information theory. These statements
are no quotations, unless indicated explicitly.

 
\begin{itemize}
\item[]{\em Einstein/Podolsky/Rosen:} An entangled wavefunction 
   does not
 describe the physical reality in a complete way.
\item[]{\em  E. Schr\"odinger:} For an entangled state
``the best possible knowledge of the whole
  does {\em not} include the best possible knowledge of its parts.''
  \cite{es}
  \end{itemize}
  Entanglement is...
  \begin{itemize}
\item[]{\em  J. Bell:} ...a correlation that is stronger than any classical
    correlation.
\item[]{\em  D. Mermin:} ...a correlation that contradicts 
the theory of elements of reality.
\item[]{\em  A. Peres:} ``...a trick that quantum magicians use to produce 
phenomena
that cannot be imitated by classical magicians.''
\cite{asherpriv}
\item[]{\em  C. Bennett:} ...a resource that enables quantum teleportation.
\item[]{\em  P. Shor:} ...a global structure of the wavefunction that allows
     for faster algorithms.
\item[]{\em  A. Ekert:} ...a tool for secure communication.
\item[]{{\em   Horodecki} family:} ...the need for first applications
      of positive maps in physics.
\end{itemize}
Our view of the nature of entanglement may continue to
be modified during the coming years. 

\section{   Given a quantum state, 
         is it {\em separable} or 
         {\em entangled}? }
  \label{seporent}
 In this section we will summarize {\em operational} 
 and {\em non-operational} criteria
 that allow to classify a given state as separable or entangled.  
 Here the word ``operational'' is  used in the sense of ``user-friendly'':
 an operational criterion is  a recipe that can be applied to an
 explicite  density matrix $\varrho$, giving some immediate answer like
     ``$\varrho$ is entangled'', or  ``$\varrho$ is separable''
     or ``this criterion is not strong enough to decide
     whether $\varrho$ is separable or entangled''.
     
 But, first of all, we need a mathematical definition for entanglement
 versus separability. This is very simple for {\em pure states}:
  a pure state $\ket{\psi}$ is called {\em separable}
   iff it can be written as
 $\ket{\psi}=\ket{a}\otimes\ket{b}$, otherwise it is {\em entangled}.
 (Remember that throughout most of this article we talk about bipartite
 entanglement. In some cases the generalisation to more particles is
 straightforward, like here.)
 An example for a pure separable state is  
$
\ket{\psi}=\ket{00}
$,
examples  for pure entangled states are the {\em Bell states}
\bea
\ket{\Phi^\pm}&=&\shalf (\ket{00}\pm \ket{11})\ ,
\label{phi}\\
\ket{\Psi^\pm}&=&\shalf (\ket{01}\pm \ket{10})\ .
\label{psi}
\eea

A {\em mixed state} is called separable, if it can be prepared 
by the two parties (which are traditionally called Alice and Bob)
in a ``classical'' way, that is, by agreeing over the phone
on the {\em local} preparation of states. A density matrix that 
has been created in this way can only contain classical correlations.
Mathematically this means: 
 a mixed state $\varrho$ is called {\em separable}
   iff it can be written as \cite{wer}
 \be
 \varrho \, { =}\, \sum_ip_i\proj{a_i}\otimes\proj{b_i}\ ,
 \label{sep}
 \ee 
 otherwise it is {\em entangled}. Here the coefficients $p_i$ are
 probabilities, i.e. $0\leq p_i \leq 1$ and $\sum_i p_i=1$.
 Note that in general $\bra{a_i}a_j\rangle \neq \delta_{ij}$,
 and also Bob's states need not be orthogonal. 
 This decomposition is not unique.
 An example for a mixed 
 separable state that contains classical correlations,
 but no quantum correlations, is
 $
\varrho = 
\half(\proj{00}+\proj{11})
$.
An 
example  for a mixed entangled state is a Werner state,
an admixture of a Bell state as in (\ref{phi}) or (\ref{psi})
to the identity:
$
\varrho_W = (1-p)\quarter\eins+p\proj{\Phi^+}$  with
 $1/3< p\leq 1$. The lower limit of $p$ for $\varrho_W$
 to be entangled can be easily found with the operational criteria 
 discussed below. In fact, any density matrix that is ``close
 enough'' to the identity is separable \cite{karol}.
 
 Finding a decomposition  as in (\ref{sep}) for a given $\varrho$,
or  proving that it does not exist,
 is a non-trivial task which has been  solved explicitly only for
a few cases.
 Therefore this simple-looking
definition of separability is by no means ``user-friendly'', and
we are in demand of criteria that are easier to test.

\subsection{Operational separability criteria}
Here  we present some
separablity criteria that are easy to check in an explicit case.
In the following we will assume that 
$
\varrho \in \calh_A\otimes \calh_B$
with dim $\calh_A=M $ and dim $\calh_B=N\geq M$, without loss
of generality.

For {\em pure states} there is a very simple necessary and
sufficient criterion for separability, the {\bf Schmidt decomposition}.
A pure state has Schmidt rank $r\leq M$ if it can be decomposed as
the bi-orthogonal sum
\be
\ket{\psi^r}=\sum_{i=1}^r a_i\ket{e_i}\ket{f_i}\ ,
\label{schmidt}
\ee
with 
       $a_i>0$ and $ \sum_i^ra_i^2=1$,
where  $\braket{e_i}{e_j}=\delta_{ij}=\braket{f_i}{f_j}$.
Note that $a_i^2$ are the eigenvalues of the reduced density
matrices, and therefore the Schmidt rank is easy to compute.
A given pure state 
 $ \ket{\psi}$ is separable iff $r=1$.

For 
{\em mixed states} the situation is less simple. There are
several operational separability criteria for this case.
Here they are ordered in decreasing strength, i.e. the last criterion 
fails to detect an entangled state as entangled in more cases
than the previous ones:
 \begin{itemize}
 \item[1)] \ {\bf  Peres-Horodecki criterion (positive partial transpose)}
 \cite{peres,ppt}: \newline 
 The partial transpose of a composite density matrix is
 given by  transposing  only one of the subsystems.
 Thus, the entries of a density matrix that is
 partially transposed with respect to Alice  are given by
 \be
 (\varrho^{T_A})_{m\mu ,n\nu } = \varrho_{n\mu ,m\nu }\ ,
 \ee
 where Latin indices are referring to Alice's subsystem and
 Greek ones to Bob's subsystem. As any separable state can be
 decomposed according to  (\ref{sep}), its partial transpose
  is  given by
 \be
 \varrho^{T_A}_{sep} = 
\sum_i p_i \, (\proj{a_i})^T\otimes\proj{b_i}\ .
\ee
Since the $(\proj{a_i})^T$ are again  valid density matrices for Alice,
one finds immediately that $\varrho^{T_A}_{sep}\geq 0$. 
The same holds for partial transposition with respect to Bob (or
any other party for multipartite systems).
In conclusion,  
the partial transpose of a separable state $\varrho$
with respect to any subsystem is positive \cite{peres}.
(In our terminology a positive operator has positive or vanishing
eigenvalues -- more precisely it should be called 
positive semidefinite. The expectation value of a positive operator
with any state is positive or zero.)

It was shown in \cite{ppt} for bipartite systems
that the converse 
(i. e. if $\varrho^{T_A}\geq 0$ then $\varrho$ is separable)
is true only for 
low-dimensional systems, namely for composite states of dimension
$2\times 2$ and $2\times 3$. In this case the positivity of the partial
transpose (PPT) is a necessary and sufficient condition for separability.
For higher dimensions it is only necessary, and the existence of 
entangled PPT states
 has been shown \cite{pawel} -- these states have been
called {\em bound entangled states}, as their entanglement does not seem 
to be ``useful'', as explained in section \ref{useful}.

\item[2)]\  {\bf   Reduction criterion} \cite{reduction}: \newline
According to the reduction criterion,   
if $\varrho$ is separable then
\be
\varrho_A\otimes \eins - \varrho \geq 0\ \ \  \mytext{and} \ \ \
\eins\otimes  \varrho_B- \varrho \geq 0\ ,
\label{red}
\ee
where $\varrho_A$ is Alice's reduced density matrix, and 
$\varrho_B$  Bob's. In order to understand why the positivity
of the left hand sides in (\ref{red}) is a   
separability criterion, one has to note that they correspond to the
application of the positive map 
 $\Lambda(\sigma)=(\mytext{Tr}\sigma)\eins - \sigma$   to 
Bob's subsystem,  or to
Alice's subsystem. (The important role of positive maps
will be discussed in the next subsection, \ref{nonop}.) 
A positive map 
applied to one subsystem of a separable state
preserves the properties of a density
matrix -- therefore the resulting density matrix  has to remain positive.

Like the partial transpose criterion, 
the reduction criterion is a
 necessary and  sufficient separability condition  only
 for dimensions $2\times 2$ and $ 2\times 3$, and
  a  necessary   condition otherwise.
  
\item[3)] \ {\bf   Majorization criterion} \cite{major}: \newline
The majorization criterion says that if a state $\varrho$ is
separable, then
\be
 {\lambda}^{\downarrow}_{\varrho}\prec {\lambda}^{\downarrow}_{\varrho_A}\ \ \
 \mytext{and} \ \ \ {\lambda}^{\downarrow}_{\varrho}\prec 
 {\lambda}^{\downarrow}_{\varrho_B}
 \label{maj}
\ee
has to be fulfilled. Here
 $\lambda^{\downarrow}_{\varrho}$ denotes the vector consisting
 of the eigenvalues
 of $\varrho$,  in decreasing order, and
a vector $ x^{\downarrow}$ is {\em majorized} by a vector $y^{\downarrow}$,
denoted as 
$ x^{\downarrow}\prec  y^{\downarrow}$,  when
 $\sum_{j=1}^k x_j^{\downarrow} \leq \sum_{j=1}^k y_j^{\downarrow}$ 
 holds  for
 $k=1,
...,d-1$, and the equality
holds for $k=d$, with $d$ being the dimension of  the vector.
Zeros are appended to the vectors ${\lambda}^{\downarrow}_{\varrho_{A,B}}$
in (\ref{maj}), in order 
to make their dimension equal to the one of  $\lambda^{\downarrow}_{\varrho}$.

Thus, for a separable state the ordered vector of eigenvalues
for the whole density matrix is majorized by the ones of the reduced
density matrices. This is was summarized by Nielsen and
Kempe as 
``Separable states are more disordered globally than locally''
\cite{major}.
Note that the spectra of a density matrix and its 
reduced density matrices
 do not allow to distinguish separable
and entangled states. The majorization criterion is only a necessary,
not a sufficient
condition for separability. 
\end{itemize}
 
The logical ordering
of the  separability criteria introduced in this section is
as follows:
\begin{itemize}
\item[-]{\em dimension $2\times 2$ and $ 2\times 3$:} \newline
$\varrho$ \, is separable  $\Leftrightarrow$ satisfies PPT
$\Leftrightarrow$ satisfies reduction criterion ${\Rightarrow}$
satisfies majorization criterion 
\item[-]{\em higher dimensions:} \newline
$\varrho$ \, is separable ${\Rightarrow}$ satisfies PPT ${\Rightarrow}$
satisfies reduction criterion $\stackrel{?}{\Rightarrow}$ 
satisfies majorization criterion  \newline
\end{itemize}

\subsection{Non-operational separability criteria}
\label{nonop}

In this subsection we will discuss two non-operational 
separability criteria. Both are necessary and sufficient
criteria for any bipartite system. They bear the major
problem, however, that they do not provide us with a simple procedure 
to check the separability properties of a given
state. This will become clear from their description.

\begin{itemize}
\item[1)] \ {\bf  Positive maps:}\newline
It was shown in \cite{ppt} that 
       {\em $\varrho$ is separable iff 
       for {\em any} positive map $\Lambda$  
     \be
     (\eins\otimes \Lambda ) \varrho \geq 0
     \label{posmap}
     \ee
holds.}

A positive map is a map that takes positive  operators
to positive operators.
 A positive map $\Lambda$
is called {\em completely positive} (CP),
if any extension  to a larger Hilbert space, i.e.
$\eins_x\otimes \Lambda $, is a positive map. Here $x$ denotes
the dimension of the extension and is arbitrary. It is clear
from equation (\ref{posmap}) that for the purpose of finding
separability criteria only those maps are interesting which are
positive, but {\em not} CP, as a CP map will fulfill (\ref{posmap}) for
any given $\varrho$. 

In the previous section we have already studied two examples
for positive maps that are not CP and the extensions of
which provide separability criteria:
the transpose  and the map 
$\Lambda(\sigma)=(\mytext{Tr}\sigma)\eins - \sigma$.
There we have already explained the reason why 
(\ref{posmap}) has to hold for separable states: it is
the possibility to 
decompose a separable state
  into a sum of tensor products according to (\ref{sep}).
  Applying a positive map to one of the subsystems will keep
  each term positive, and therefore also their sum.
  
Note that the problem about  the non-operational criterion
of positive maps lies in the little word ``any'' just before
(\ref{posmap}): we do not have a complete characterization
of the set of all positive maps.

\item[2)] \ {\bf   Entanglement witnesses:}
\newline
The criterion of the so-called entanglement witnesses was
given in \cite{ppt} and studied in \cite{witness}:

{\em A density matrix $\varrho$ is entangled iff 
there exists a Hermitian operator $\calw$ 
    with 
  Tr$(\calw \varrho) < 0$ and 
    Tr$(\calw \varrho_{sep}) \geq  0$
  for any separable state $\varrho_{sep}$.}
  
We say that  the witness $\calw$ ``detects'' the entanglement of $\varrho$. 

\item[]
{\bf  Correspondence between 1) and 2):}
\newline
These two criteria are not independent -- there is the
 Jamio\l kowski isomorphism \cite{jami} that provides us
with a correspondence between an entanglement witness and a positive
map,
\be
\calw = (\eins\otimes \Lambda ) P_+\ ,
\ee 
where $P_+ = \frac{1}{M}(\sum_{i=1}^M\ket{ii} )
(\sum_{j=1}^M\bra{jj} )$ is the projector onto the maximally entangled
state.
\end{itemize}

Let us for the rest of this subsection pursue the concept of
entanglement witnesses. What seems at first sight to be a rather 
abstract theorem will prove to be a powerful tool,
as it allows to answer explicit questions about  entanglement properties 
of certain states, see e.g. subsection \ref{trip}.

The existence of entanglement witnesses
is a  consequence of  the Hahn-Banach theorem, which states
the following:  \newline
 {\em Let $S$ be a convex, compact set, and let   
 $\varrho \not\in S$. Then  there exists a hyper-plane that
 separates $\varrho$ from $S$.} 
 
\begin{figure}[ht]
\vspace*{-2.5cm}
\hspace*{4cm}
\setlength{\unitlength}{1pt}
\begin{picture}(500,300)
\psfig{file=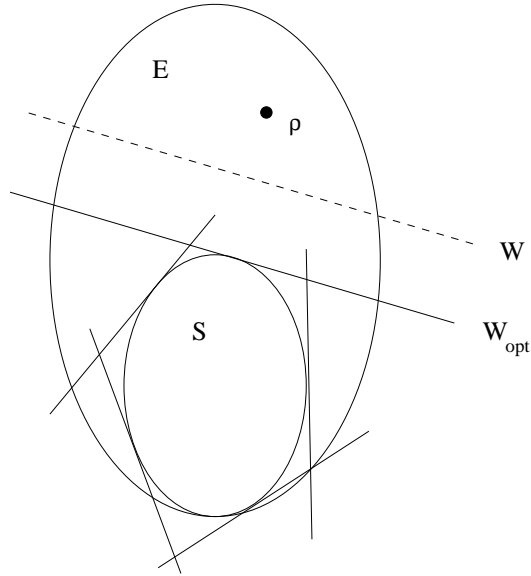,width=7cm}
\end{picture}
\vspace*{0.2cm}
\caption[]{\small Intuitive picture of entanglement witnesses
and their optimization.}
\label{wit}
\end{figure}

 This fact is illustrated in figure 
 \ref{wit}: the bigger ``egg''-shaped  set symbolizes the
 convex, compact set of all density matrices. The smaller one stands
 for the separable density matrices, and is a convex, compact subset
 of the bigger one. The state $\varrho$ is entangled and therefore
 $\not\in S$. The dotted line sketches the hyper-plane that separates
 $\varrho$ from $S$, and is given by those $\sigma$ that fulfill
 Tr$(\calw \sigma) = 0$. 
 
 It is helpful 
 to realize that  Tr$(\calw \varrho) $ defines a scalar product. Let us for a 
 moment look at a scalar product  which is  familiar to everybody,
 the  scalar product of two vectors of unit length
 with common origin, let us call them
 $\vec w $ and $\vec r$. Their scalar product $\vec w \cdot \vec r$
 is equal to $\cos \alpha$, where $\alpha$ is the relative angle
 between the two vectors. 
 A {\em fixed } $\vec w $ defines a 
 certain plane -- the one to which $\vec w $
 is orthogonal. Vectors $\vec r$ from this plane have a vanishing
 scalar product with $\vec w$.
 All vectors $\vec r$ that are ``on one side'' of this
 plane  have a positive scalar product with $\vec w $, all
 vectors $\vec r$ ``on the other side'' 
  a negative one -- due to the properties of the cosine function.
 The scalar product Tr$(\calw \varrho) $ has the same property:
 all density operators on one side of the hyper-plane lead to a
 positive outcome, the ones on the other side to a negative
 one. 
 
 This intuitive picture of entanglement witnesses also helps to
 understand how they can be optimized \cite{witopt}: performing
 a parallel transport of the hyper-plane such that it becomes
 tangent
 to the set of separable states means that the 
 corresponding optimized
 witness $\calw_{opt}$ detects more entangled states
 than before. This is also
 indicated in figure \ref{wit}. 
 
 In order to completely characterise the set $S$ one would in principle
 need infinitely many witnesses, unless the shape of $S$ is a 
 polytope. This is not known nowadays. But several witnesses
 can already give a good approximation of the set of separable states.
 Methods to construct entanglement  witnesses in a canonical way have been
 provided in \cite{witopt}.
         
\section{Given an entangled state, 
         is the entanglement 
         {\em useful}?}
\label{useful}
     
 In an ideal experiment an initially prepared maximally entangled
 state would remain maximally entangled. In reality, the resource
 of entanglement is  very fragile, due to interaction with 
 the environment. As entanglement is the foundation of many
 quantum information processing tasks, it would
 therefore be desirable to concentrate non-maximal entanglement.
 A central question is: given several copies of a 
 non-maximally  entangled state, is there a process
 that allows to locally ``distill'' its entanglement, i.e. to retrieve
 a maximally entangled state?
      
\begin{figure}[ht]
\vspace*{1.5cm}
\hspace*{4cm}
\setlength{\unitlength}{1pt}
\begin{picture}(500,300)
\put(1,320){a)}
\put(1,220){b)}
\put(1,140){c)}
\put(1,30){d)}
\psfig{file=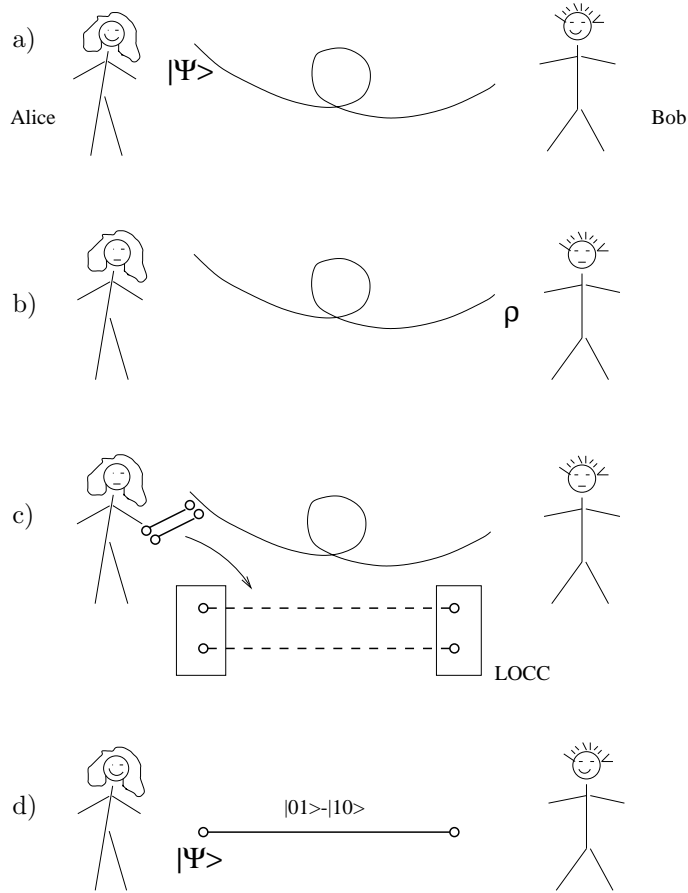,width=9cm}
\end{picture}
\vspace*{0.2cm}
\caption[]{\small Providing a noiseless channel via distillation:
a) Alice wants to send the message $\ket{\psi}
$ to Bob.
b) Bob receives $\varrho$ instead, as the channel is noisy.
c) Alice sends one subsystem of a maximally entangled state
   through the noisy channel to Bob, and repeats this with a second
   pair. They employ a distillation protocol. 
d) Alice and Bob have created a maximally entangled singlet
   which they can use as a noiseless  teleportation channel.}
\label{distill}
\end{figure}

 This is a straightforward motivation to study entanglement 
 distillation. In addition, this concept  is useful in any other
 quantum communication task: assume that Alice wants to send 
 a quantum message to Bob. As a  simple illustration   she will send
 a polarized photon along an optical fiber.       
 Interaction with the environment disturbs 
 the original quantum state -- in general one has to deal with
 ``noisy channels''. An initial pure state $\ket{\psi}$ will 
 arrive as  some disturbed mixed state $\varrho$. This 
 situation is sketched in figure 
 \ref{distill} a) and b). 
 
 In the wide field of 
 error correction one deals with this problem   
  by ``repairing the state'', i.e. via
  encoding, finding the error and then restoring the original state.
 The idea of  distillation of non-maximally entangled states
 pursues a different path, by  
 ``providing a noiseless channel'', as explained below.

 Figure \ref{distill} c) visualizes this concept: 
Alice sends one subsystem of a maximally entangled state
   through the noisy channel to Bob. The resulting state
   will not be maximally entangled and mixed, due to the noise. 
   She  repeats this with a second pair or more
   pairs. Alice and Bob then operate locally 
   on their respective qubits and communicate classically
   (LOCC= local operations and classical communication),
   thus 
    employing a distillation protocol. One explicit protocol
    will be introduced below. Thus they  create a maximally
    entangled state as indicated in figure \ref{distill} d). 
    This state can now be used as a noiseless 
    channel via teleportation.
    
  In summary, it is an essential question to ask:
 Given an entangled density matrix
  $\varrho$, can  its entanglement be {\em distilled}?
  
 \subsection{A distillation protocol}
  The following distillation protocol 
  for non-maximally entangled mixed states was proposed in
  \cite{distprot}. It is designed for the case that
  Alice and Bob share  a supply
   of many identical entangled 
   bipartite
  systems of qubits. They can always  
      convert them by local operations  to the isotropic state
      $\varrho_{iso}=(1-p)\frac{1}{4}\eins
         +p \proj{\Phi^+}$ ,  with  $1/3< p\leq 1$.
  
\begin{figure}[h]
\vspace*{-7.cm}
\hspace*{4cm}
\setlength{\unitlength}{1pt}
\begin{picture}(500,300)
\psfig{file=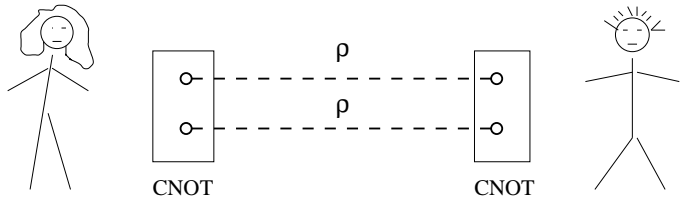,width=9cm}
\end{picture}
\vspace*{0.2cm}
\caption[]{\small First step in the distillation protocol:
    applying local CNOT's.}
\label{firststep}
\end{figure}

  In the first step     
 Alice and Bob use two $\varrho$'s, 
 as illustrated in figure \ref{firststep}.
 Each of them applies a
 local CNOT-gate to his/her two qubits. The action of this gate is given by
   $U_{CNOT} \ket{a_1}\ket{a_2}=\ket{a_1}\ket{(a_1+a_2) 
    \mytext{mod}\, 2}$.
    
\begin{figure}[h]
\vspace*{-7cm}
\hspace*{4cm}
\setlength{\unitlength}{1pt}
\begin{picture}(500,300)
\psfig{file=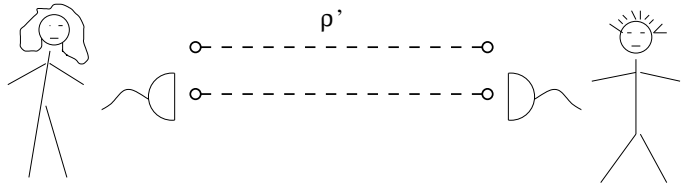,width=9cm}
\end{picture}
\vspace*{0.2cm}
\caption[]{\small Second step in the distillation protocol:
   measurement.}
\label{secstep}
\end{figure}

 In the next step, both Alice and Bob do a measurement on their
 second qubit, as shown in figure \ref{secstep}. They only
 keep the first density matrix, which had
  changed to some $\varrho'$, if their outcomes are
 identical. Otherwise the two pairs have to be discarded.

\begin{figure}[h]
\vspace*{-3.5cm}
\hspace*{4cm}
\setlength{\unitlength}{1pt}
\begin{picture}(500,300)
\psfig{file=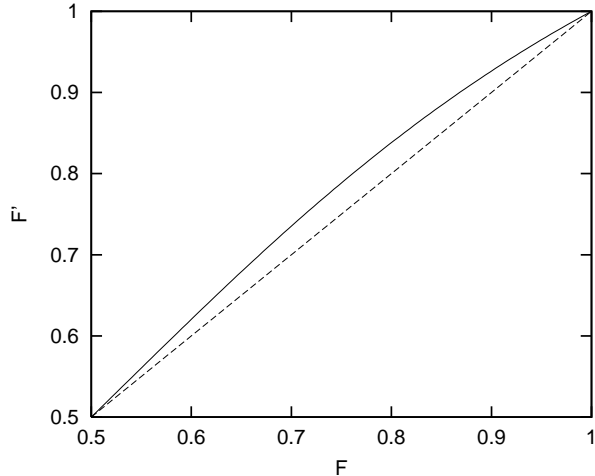,width=8cm}
\end{picture}
\vspace*{0.2cm}
\caption[]{\small New fidelity after one distillation step, as
  function of the fidelity in the previous step.}
\label{fidel}
\end{figure}

 This process has increased the overlap of the new density matrix
 $\varrho'$
 with the maximally entangled state. 
 Thus both entanglement and purity are enhanced.
 The new fidelity is defined as
  $F'=\bra{\Phi^+}\varrho'\ket{\Phi^+}$ and is plotted in figure
  \ref{fidel}, as function of the original fidelity
  $F=\bra{\Phi^+}\varrho\ket{\Phi^+}=(1+3p)/4$. This procedure  is
   then repeated with new pairs of the higher fidelity. In this way the
  entanglement is increased  in successive steps, finally being maximal
  when enough original pairs were available.
  A distillation protocol is successful, i.e.
  enhances the entanglement, whenever
  the curve for the new fidelity lies above the line
  $F'=F$ (dashed line in figure \ref{fidel}).
  
 \subsection{Which states
   can   be {\em distilled}?}
   For the general reasons discussed at the beginning of 
   section \ref{useful} it is a very fundamental question to ask: which
   entangled states can be distilled?
   For two-qubit states the answer was given in \cite{horodist}:
 {\em all} entangled two-qubit states 
 are distillable.
 
 In general, this question is unsolved, however. A necessary
 and sufficient criterion for distillability of a given
 $\varrho$ was proved in \cite{horobe}:
 
 {\em The state
 $\varrho$ is  distillable iff there exists 
 $\ket{\psi^2}=a_1\ket{e_1}\ket{f_1}+a_2\ket{e_2}\ket{f_2}$ 
  such that 
 $\bra{\psi^2}(\varrho^{T_A})^{\otimes n}\ket{\psi^2} < 0$ for some $n$.}
 
 In other words, if for a certain number $n$ of copies the
 partial transpose of the total state has a negative expectation value
 with some vector of Schmidt rank 2, then $\varrho$ can be distilled
 (one says: $\varrho$ is $n$-distillable), and
 vice versa. From this theorem it follows immediately that a state
 with a positive partial transpose cannot be distilled: if 
    $\varrho^{T_A}\geq 0$,
    then $(\varrho^{T_A})^{\otimes n}\geq 0$, and thus
    PPT-states are undistillable. As mentioned above, 
    entangled PPT-states are therefore
    called ``bound entangled''.

It is an   open question whether the reverse of the statement
``PPT-states are undistillable''
is also true, i.e. if a state is undistillable, does it have to be
PPT? For the dimension 
 $2\times N$ this  is indeed true \cite{npt1}. For higher dimensions
 there is a strong conjecture that it is false, i.e. that there
 are undistillable states with a non-positive partial transpose
 (NPT). A family of such states in dimension $N\times N$
 was discussed in
 \cite{npt1,npt2}. This family consists of a convex combination of 
 projectors onto 
 the symmetric and 
 the antisymmetric subspace,
 where the relative weight of these two contributions
 is the only free parameter. Depending
 on the value of this parameter, the state is PPT or NPT. 
 The undistillability of  NPT states with a 
 certain constant finite range of this parameter
 was shown numerically for up to 3 copies in the case $N=3$.
 The problem in finding a rigorous answer to the
 question of distillability for these NPT states
  lies in the fact that one has
 to consider the limit $n\rightarrow \infty$.

\begin{figure}[h]
\vspace*{-4.5cm}
\hspace*{4cm}
\setlength{\unitlength}{1pt}
\begin{picture}(500,300)
\psfig{file=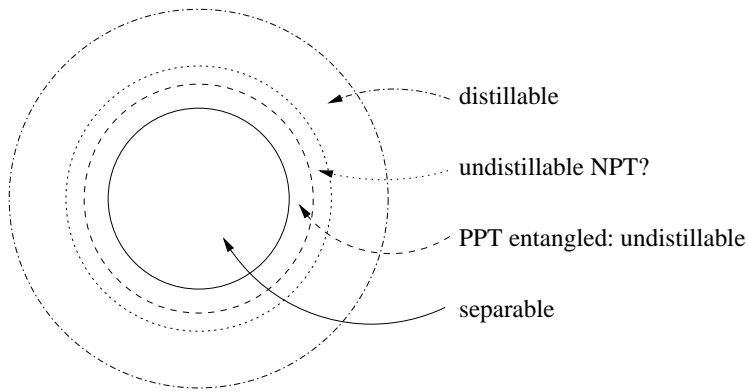,width=10cm}
\end{picture}
\vspace*{0.2cm}
\caption[]{\small Decomposition of the set of all states
into distillable and undistillable states.}
\label{setdist}
\end{figure}

Our present understanding of how the set of all states is decomposed
into separable, entangled undistillable and distillable states
is summarized in figure \ref{setdist}.

\section{
Given an entangled state, 
 {\em how much} is it entangled?}
\label{howmuch}
 
So far we have shown that it is still an
open question how to {\em qualify} a given state as 
separable versus entangled or undistillable versus distillable.
How can we hope to answer the question of {\em quantifying}
the amount of entanglement of a given state? It is not surprising
that there is no simple answer to that.
We will  summarize the requirements for
a good entanglement measure, and introduce the reader to
some important entanglement measures, without making the
 attempt to discuss all
existing entanglement measures.

In  subsection \ref{schmidtsec}
we will explain the concept of Schmidt witnesses, which 
allows to
classify  entangled states in classes according to their
Schmidt number. In the final two subsections
we will study composite systems of three qubits, showing that
the question of quantification of entanglement has to be
reformulated in that case.

\subsection{ Requirements for entanglement measures } 
 
 A good entanglement measure $E$ has to fulfill several
 requirements. 
 However, it is still an open question
  whether all of these conditions are indeed
 necessary. 
 In fact, some of
 the entanglement measures that are introduced below do not
 fulfill the whole list of properties, see Table I. 
 
 \begin{itemize}
 \item[1)] If $\varrho$ is {\em separable} then  $E(\varrho)=0$.
 \item[2)]\ {\em Normalization:} the entanglement of a maximally entangled
     state  of two $d$-dimensional systems is given by
 \be E(P_+^d)=\log d\ .\ee
  \item[3)]\ {\em No increase under LOCC:}  applying local
  operations to $\varrho$ and classically communicating cannot
  increase the entanglement of $\varrho$, i.e. 
      \be E(\Lambda_{LOCC}(\varrho)) \leq E(\varrho)\ .\ee 
 \item[4)]\ {\em Continuity:} In the limit of vanishing distance
  between two density matrices 
     the difference between  their  entanglement should tend to zero, i.e.
 \be E(\varrho)-E(\sigma) \rightarrow 0 \ \ \ \mytext{for} \ \ \ 
  ||\varrho - \sigma || \rightarrow 0\ .\ee 
\item[5)]\ {\em Additivity:} A certain number $n$ of 
identical copies
of the state $\varrho$ should contain $n$ times the entanglement of 
 one copy,
 \be E(\varrho^{\otimes n})=n\, E(\varrho)\ .\ee
 \item[6)]\ {\em  Subadditivity:} The entanglement of 
 the tensor product of two states
 $\varrho$ and $\sigma$ should not be larger than the sum of 
 the entanglement of each of the states,
 \be E(\varrho \otimes \sigma ) \leq E(\varrho )+E(\sigma)\ .\ee
 \item[7)]\ {\em  Convexity:}  The entanglement measure should be 
 a convex function, i.e.
 \be E(\lambda \varrho +(1-\lambda)\sigma ) \leq 
 \lambda E(\varrho)+(1-\lambda)E(\sigma)\ee
 for $0< \lambda < 1$.
 \end{itemize}
 
 \subsection{Some important entanglement measures }
 For a pure bipartite state $\ket{\psi}$  a good entanglement measure
 is the von Neumann entropy of its reduced density
 matrix, $S(\varrho_{red}) = -\mytext{Tr}(\varrho_{red}
  \log \varrho_{red})$.
  For mixed states there is no unique entanglement measure,
  but all entanglement measures should coincide on pure
  bipartite states and be equal to the von Neumann entropy
  of the reduced density matrix (uniqueness theorem)
  \cite{mdonald}.
 Some important entanglement measures are defined as follows:

 \begin{itemize}
 \item[-]
  {\em Entanglement cost:} 
 The entanglement cost tells us how expensive it is to create
 an entangled state $\varrho$, i.e. what is the ratio of
 the number of
 maximally entangled input states $\ket{\Phi^+}$
 over the produced output
 states $\varrho$, minimized over all LOCC operations. In the 
 limit of infinitely many outputs this reads
 \be
   E_C(\varrho) = \inf_{\{\Lambda_{LOCC}\}}
                     \lim_{n_\varrho\rightarrow \infty}
                     \frac{n^{in}_{\ket{\Phi^+}}}{n^{out}_\varrho}\ .
   \ee
 \item[-]  
  {\em Entanglement of formation:} 
 Any state $\varrho$ can be decomposed as a convex combination  of
 projectors onto 
 pure states, $\varrho =\sum_ip_i\proj{\psi_i}$.
 The entanglement of formation is the averaged von Neumann
 entropy of the reduced density matrices of the 
 pure states $\ket{\psi_i}$,
 minimized over all possible decompositions,
 \be
   E_F(\varrho) =\inf_{\{dec\}}\sum_ip_iS(\varrho_{i,red})\ .
   \ee
 \item[-]  
  {\em Relative entropy of entanglement:} 
 The relative entropy can be seen intuitively as the ``distance'' of
 the entangled $\varrho$ to the closest separable state $\sigma$,
 although it is not a  distance in the mathematical sense,
 \be
   E_R(\varrho) =\inf_{\sigma\in S}\mytext{tr}[\varrho(\log \varrho 
   -\log \sigma)]\ .
   \ee
 \item[-]  
  {\em Distillable entanglement:} 
 The distillable entanglement tells us how much entanglement
 we can extract from 
 an entangled state $\varrho$, i.e. what is the ratio of
 the number of
 maximally entangled output states $\ket{\Phi^+}$ over the needed input
 states $\varrho$, maximized over all LOCC operations. In the 
 limit of infinitely many inputs this reads
   \be
   E_D(\varrho) = \sup_{\{\Lambda_{LOCC}\}}
            \lim_{n_\varrho\rightarrow \infty}\frac{n^{out}_{\ket{\Phi^+}}}
                     {n^{in}_\varrho}\ .
   \ee
 \end{itemize}
 
 There are some known 
 relations between these entanglement measures: the distillable entanglement
 (entanglement cost)
 is a lower (upper) bound for {\em any } entanglement measure, i.e.
 $ E_D(\varrho)\leq  E(\varrho) \leq  E_C(\varrho)$.
 For any bound entangled state $ E_D(\varrho) < E_C(\varrho)$
 holds, but there is also an example for a free entangled state,
 i.e. a distillable state, with the same property \cite{vidcir}.
 It is conjectured that the entanglement of formation and the entanglement 
 cost are identical, i.e. 
 $E_F(\varrho)\stackrel{?}{=}E_C(\varrho)$.
 
 Some known and unknown properties of the entanglement measures 
 discussed above 
 are given in Table I.
 
 \vspace*{0.7cm}
 
 \begin{table}[h]
 \begin{tabular}{l|cccc}
  & $\ \ E_C\ \ $ & $\ \ E_F\ \ $ & $ \ \ E_R\ \ $ & $\ \ E_D\ \ $ \\
                        \hline
    \mytext{continuity} & ? & $\surd$ & $\surd$& ?\\
    \mytext{additivity} & $\surd$ & ? & \ \ no\cite{werner} & $\surd$
    \\
    \mytext{convexity} & $\surd$ & $\surd$ & $\surd$& \ \ no (?)\cite{ibm}
 \end{tabular}
 \vspace*{0.5cm}
 \caption[]{\small Properties of entanglement measures.}
 \end{table}
 
\subsection{Schmidt witnesses}
   \label{schmidtsec}
   A slightly different question from ``how much entangled is a state
   $\varrho$?''
   can be addressed via the generalization of entanglement witnesses
   to so-called Schmidt-witnesses. They give an answer to the
   question ``how many degrees of freedom are
     entangled in $\varrho$?'' This corresponds to a  finer
   classification of entangled states.

The Schmidt rank for pure bipartite states,
as defined in in equation (\ref{schmidt}), was 
generalized to the so-called 
 Schmidt number $k$ for mixed bipartite states in \cite{barpaw}.
 For a given decomposition of $\varrho$ into a convex combination  
 of projectors onto pure states, 
 let us call the highest occurring Schmidt rank 
 $r_{\rm{max}}$. The Schmidt number is the minimization 
 of this highest Schmidt rank over
 all possible decompositions:
\be
\varrho = \sum_i p_i\ket{\Psi_i^{r_i}}\bra{\Psi_i^{r_i}} \  , \ \ \ 
k=\min_{\{ dec\} }( r_{\rm{max}})\ .
\ee
The Schmidt number cannot be higher than $M$,
the smaller of the  dimensions of the two subsystems.

\begin{figure}[h]
\vspace*{-2.5cm}
\hspace*{4cm}
\setlength{\unitlength}{1pt}
\begin{picture}(500,300)
\psfig{file=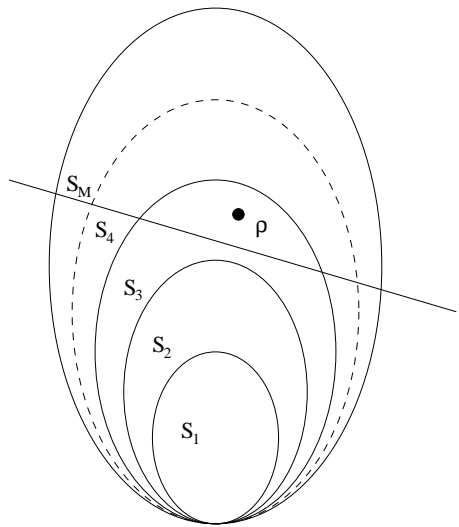,width=6cm}
\end{picture}
\vspace*{0.2cm}
\caption[]{\small Schmidt classes and the detection of the
Schmidt number by a Schmidt witness.}
\label{schmidtfig}
\end{figure}

Entangled states can now be classified according to
 Schmidt classes \cite{adm}: the Schmidt class
$S_k$ is a subset of the set of all density matrices and 
contains all density matrices with Schmidt number ${\leq k}$.
The Schmidt classes are successively embedded into each other, as
visualized in figure \ref{schmidtfig}: 
$S_1 \subset S_2 \subset ... \subset S_M$. 

A Schmidt witness $\calw_k$ for the Schmidt class $S_k$
is defined as a straightforward generalization of  the entanglement
witnesses discussed in subsection \ref{nonop}. 
A Schmidt witness $\calw_k$ is a Hermitian operator for which
 \bea
 \exists \, \varrho \in S_{k} \mytext{ with } \ \ 
  & \ \ \ \mytext{Tr}(\calw_k \varrho) \ \ < 0& \ ,\\
 \ \  \forall \varrho_{k-1} \in S_{k-1}:\ \ 
&\mytext{Tr}(\calw_k \varrho_{k-1}) \geq  0&\ .
 \eea
 
 Constructive methods for 
      Schmidt witnesses have been shown; they can be
optimized  in analogy with entanglement witnesses,
and they can be used for example as a  tool to study  
properties of bound entangled
states.

\subsection{``Many'' systems: pure three-qubit states}
\label{many}
So far we have only considered bipartite systems.
Unfortunately, according to nowadays'
knowledge one already has to call three subsystems ``many''. 
Can we generalise the concepts that were introduced
so far to tripartite states? Again, one can ask whether a
given state is separable or entangled. But now one can also
specify  the kind of entanglement: is it genuine three-particle
entanglement or are just two of the three subsystems entangled? 
Like for bipartite states, one can ask whether a given tripartite 
state can be
distilled. This question will not be addressed here, however.
And, finally, does it make sense to ask ``how much is a given
tripartite state entangled''?

For {\em pure} three-qubit states the situation is as follows:

A
{\em separable} state  can be written as
 \be
\ket{\psi_{S}}= \ket{\phi_A}\otimes \ket{\phi_B}\otimes \ket{\phi_C}\ .
\ee

A {\em  biseparable} state is a state where only two out of the
three systems are entangled, and the third system is 
a tensor product
with the entangled ones,
 e.g. A--BC: 
\be
\ket{\psi_{B}}= \ket{\phi_A}\otimes \sum_{i=1}^2 a_i\ket{e_i}\ket{f_i}\ .
\ee
The other two possible partitions for biseparable states
are B--AC and C--AB.

A {\em three-qubit  correlated} state is one with genuine entanglement
of all three subsystems. It was shown in \cite{duer}
that  there 
exist two classes of {\em inequivalent}  states:
{
\bea
\ket{\psi_{GHZ}} &=&
\shalf (
\ket{000}+\ket{111})\ ,
\label{ghz}\\ 
\ket{\psi_{W}} &=& \sthird (\ket{100}
          +\ket{010} +\ket{001})\ .
\label{w}
\eea }
Any three-qubit correlated pure state
$\ket{\psi}$ can be transformed into { either}
$\ket{\psi_{GHZ}}$ { or} $\ket{\psi_{W}}$
by local  reversible operations ${\cal A}\otimes
{\cal B}\otimes {\cal C}$ .

Therefore, it is not enough to ask whether a given three-qubit state
is separable or biseparable or three-party entangled, but 
also whether  a genuinely three-qubit entangled
state belongs to  the $GHZ$- or $W$-class. For mixed states,
 the tool of witness operators is again useful
for this purpose, as will be discussed in the following subsection.
 
\subsection{Classification of mixed three-qubit states}
\label{trip}

Let us introduce  entanglement classes for mixed three-qubit states
\cite{abls}.
A mixed three-qubit state $\varrho$ can be written as convex combination
of pure states: if $\varrho$ can be decomposed as a sum of 
projectors onto pure separable
states $\proj{\psi_{S}}$, then it belongs to the convex compact set
$S$. If one needs at least one biseparable state $\proj{\psi_{B}}$
in the sum, but no genuine tripartite entanglement, then $\varrho$ 
belongs to the class $B$, more precisely to
$B\setminus S$. In the same way we define
the $W$-class (needs at least one W-state in the
decomposition) and the $GHZ$-class (at least one GHZ-state needed). 

\begin{figure}[ht]
\vspace*{-3.8cm}
\hspace*{2cm}
\setlength{\unitlength}{1pt}
\begin{picture}(500,300)
\psfig{file=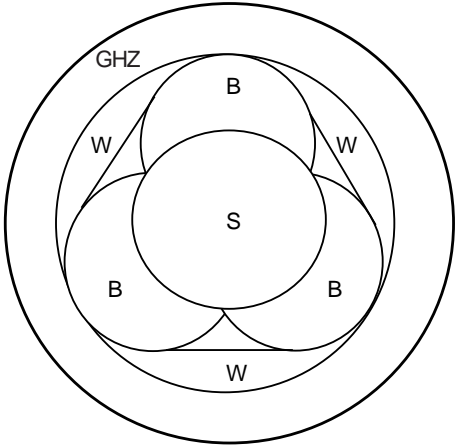,width=6cm}
\end{picture}
\vspace*{0.5cm}
\caption[]{\small  Schematic structure of the set of all three--qubit states.
        $S$: separable class; $B$: biseparable class (convex hull
of biseparable states with respect to any partition); 
         $W$--class and
        $GHZ$--class.}
\label{Fig.1}
\end{figure}

These sets are embedded into each other:
$S\subset B\subset 
W\subset GHZ$. This is schematically shown in figure \ref{Fig.1}
.
It is important to note that $W\subset GHZ$ and not the other way
round: otherwise the class $GHZ$ would not be compact, as can be seen
by studying the most general form of a W- versus a GHZ-state, 
as given in
\cite{tony}. 
 
In analogy to entanglement witnesses and Schmidt witnesses one can 
construct
  tripartite  witnesses. A 
   GHZ witness $\calw_{GHZ}$ is an Hermitian operator  with
  Tr$(\calw_{GHZ} \, \varrho) < 0$ for some
  $\varrho \in GHZ\setminus W$, and 
   Tr$(\calw_{GHZ} \, \varrho_{W}) \geq  0$ for all 
   $\varrho_{W} \in W$.
  An example for a
GHZ witness is given by 
 \be
 \calw_{GHZ} = \frac{3}{4}\one-P_{GHZ} \ ,
 \ee
 where $P_{GHZ}$ is the projector onto 
 $\ket{\psi_{GHZ}}$, given in (\ref{ghz}).
 It is straightforward to show that this
 operator  has the desired properties,
 when one realises that the maximal squared overlap between
 a pure W-state and  $\ket{\psi_{GHZ}}$ is given by 3/4.

In a similar manner one can define a
W witness. An example for such a witness that detects a W-state, but
has a positive or vanishing expectation value for all states in $B$, is
given by
\be
\calw_{W}  = \frac{2}{3}\one-P_{W}\ ,
\label{ww}
\ee
where $P_{W}$ is the projector onto 
 $\ket{\psi_{W}}$, given in (\ref{w}), and 
2/3 is the maximal squared overlap between $\ket{\psi_{W}}$ and a pure
biseparable state.

 Using the witness in (\ref{ww}), one can show that the set of mixed
W $\setminus$ B--states is   not of measure zero. This is   
 contrary to the pure case, where
 W-states are of measure zero \cite{duer}. The idea of the proof is to show
 that there is a finite ball around a state from the family
 \be 
 \varrho = \frac{1-p}{8}\one + p P_W\ ,
 \ee
 for a certain given parameter $p$, such that the ball is contained
 in the W class.
 This is an example, where  the concept of witnesses 
  helps to answer an explicit question about the structure
 of the set of entangled states.
 
 Another interesting topic in this context are the
properties of bound entangled states. Using again the tool
of witness operators, there is some 
evidence that
bound entangled three-qubit states cannot be in   GHZ $\setminus$ W,
i.e. they are at most in W \cite{abls}.

By studying mixed three-qubit states we have realized that 
in this case it is not
enough  to ask, how much a given state is entangled.
As there are different inequivalent entanglement classes,
this question makes sense only within a given class W or GHZ.
For more than three qubits 
 the number of inequivalent classes
grows fast \cite{duer}.

\section{Summary}

The section headings  in this tutorial
were phrased as questions. Let us summarize some answers.

{\em What  is entanglement?} \newline 
There are  many possible answers, maybe as many as there are
researchers in this field.

  {\em Given a state $\varrho$, is it  separable or 
         entangled}? \newline
This question is easy to answer for pure states, 
and  for low dimensions ($2\times 2$ and  $2\times 3$).
  It is very 
          difficult to answer otherwise. Several
          separability criteria have been explained.
          
 {\em Given an entangled $\varrho$, is the entanglement  useful}?
\newline
We have discussed that
  states with a positive partial transpose
  are undistillable, most  states with a non-positive partial
  transpose (NPT) are distillable; but
 some NPT states are
      conjectured to be undistillable.
      
 {\em Given an entangled $\varrho$,  how much is it entangled}?
\newline
There are  several different bipartite entanglement measures
which quantify the degree of entanglement.
 For multipartite systems there are  inequivalent 
   entanglement classes, and therefore the above question 
   has to be rephrased accordingly.

 \subsection{Acknowledgements}
Many thanks to Antonio Ac\'\i n,
 Maciej Lewenstein, Asher
Peres, Martin Plenio and Anna Sanpera for enjoyable discussions
about entanglement. 
This work has been supported  by  DFG 
(Schwerpunkt  ``Quanteninformationsverarbeitung''), 
 and the EU IST-Programme
 EQUIP.


\begin{thebibliography}{99}
\bibitem{es} E. Schr\"odinger, Naturwissenschaften {\bf 23}, 807
(1935).
\bibitem{primer} 
     M. Lewenstein, D. Bru\ss , J. I. Cirac, B. Kraus, M. Ku\'s, 
      J. Samsonowicz, A. Sanpera, and R. Tarrach,
      {\em  ``Separability and distillability in 
      composite quantum systems -- a primer''},
       Journ.  Mod. Opt. {\bf 47}, 2841 (2000).
\bibitem{hororev} 
   M.  Horodecki, P. Horodecki, and R. Horodecki,
  {\em  ``Mixed-state entanglement and quantum communication''},
    in: ``{\em Quantum Information: An Introduction to Basic
                   Theoretical Concepts and Experiments}
                 (Springer Tracts
                   in Modern Physics, 173), 
                   Eds. G. Alber, T. Beth, M. Horodecki,
         P. Horodecki, R. Horodecki,
             M. R\"otteler, H. Weinfurter, R. Werner, and
             A. Zeilinger,     Springer-Verlag (April 2001).
\bibitem{terhal}  
        B. Terhal,
        {\em  ``Detecting quantum entanglement''},
         quant-ph/0101032.
\bibitem{mdonald}  
     M. Donald, M. Horodecki, and O. Rudolph,
     {\em  ``The uniqueness theorem for entanglement
measures''},
         quant-ph/0105017.
\bibitem{qic} Quantum Information and Computation,
    Vol. 1, No. 1 (2001).
\bibitem{asherpriv} A. Peres, {\em private communication}.
\bibitem{wer} R. Werner, Phys. Rev. A {\bf 40}, 4277 (1989).
\bibitem{karol} K. \.Zyczkowski, P. Horodecki, A. Sanpera, and M. Lewenstein, 
Phys. Rev. A {\bf 58}, 883
(1998).
\bibitem{peres}  A. Peres,  
    Phys. Rev. Lett. {\bf 77}, 1413 (1996).
\bibitem{ppt}   M. Horodecki,  P. Horodecki, and  R. Horodecki,  
    Phys.  Lett. A {\bf 223}, 1 (1996).
\bibitem{pawel} P. Horodecki, Phys. Lett. A {\bf 232}, 333 (1997).
\bibitem{reduction}    M. Horodecki and P. Horodecki,  
    Phys. Rev. A {\bf 59}, 4206 (1999).
\bibitem{major}     M. Nielsen and J. Kempe,  
     quant-ph/0011117.
\bibitem{witness} B. Terhal,
    Phys. Lett. A  {\bf 271}, 319 (2000).
\bibitem{jami} A.   Jamio\l kowski,
    Rep. Math. Phys.   {\bf 3}, 275 (1972).
\bibitem{witopt} M. Lewenstein, B. Kraus, I.J. Cirac, and P. Horodecki,  
     Phys. Rev. A {\bf 62}, 052310 (2000).
\bibitem{distprot} C. Bennett, G. Brassard, S. Popescu, B. Schumacher,
 J. Smolin, and W. Wootters,  
    Phys.  Rev. Lett.  {\bf 76}, 722 (1996).
\bibitem{horodist}  M.  Horodecki, P.  Horodecki, and R. Horodecki,  
    Phys.  Rev. Lett.  {\bf 78}, 574 (1997).
\bibitem{horobe} M. Horodecki, P. Horodecki, and  R. Horodecki,  
    Phys. Rev. Lett. {\bf 80}, 5239 (1998).
\bibitem{npt1} W. D\"{u}r, J.I. Cirac, M. Lewenstein, and D. Bru\ss ,
     Phys. Rev. A {\bf 61}, 062313 (2000).
\bibitem{npt2}  D. DiVincenzo, P. Shor, J. Smolin, B. Terhal, and 
     A. Thapliyal,
     Phys. Rev. A {\bf 61}, 062312 (2000).
\bibitem{vidcir} G. Vidal and  J. I. Cirac, quant-ph/0107051.
\bibitem{werner} K. Vollbrecht and R. Werner,
     quant-ph/0010095. 
\bibitem{ibm} P. Shor, J. Smolin, and B. Terhal,
      Phys. Rev. Lett. {\bf 86}, 2681 (2001).
\bibitem{barpaw} B.~Terhal and  P.~Horodecki,
    Phys. Rev. A {\bf 61}, 040301(R) (2000).
\bibitem{adm} A.~Sanpera, D.~Bru\ss , and M.~Lewenstein, 
           Phys. Rev. A {\bf 63}, 050301(R) (2001).
\bibitem{duer} W. D\"ur, G. Vidal, and J. I. Cirac, 
     Phys. Rev. A {\bf 62}, 062314 (2000).
\bibitem{abls} A. Ac\'\i n, D. Bru\ss ,
 M. Lewenstein, and A. Sanpera, 
   Phys. Rev. Lett. {\bf 87}, 040401 (2001).
\bibitem{tony} A. Ac\'\i n,
 A. Andrianov, L. Costa, E.
Jan\'e,  J. I. Latorre, and R. Tarrach, 
Phys. Rev. Lett. {\bf 85}, 1560
(2000).
\end{thebibliography}
\end{document}